# Firm Growth and Innovation in the ERP Industry: A Systems Thinking Approach

**Srujana Pinjala**
Management Information Systems
Indian Institute of Management Calcutta
Kolkata, India
Email: pinjalas10@iimcal.ac.in

**Rahul Roy**
Management Information Systems
Indian Institute of Management Calcutta
Kolkata, India
Email: rahul@iimcal.ac.in

**Priya Seetharaman**
Management Information Systems
Indian Institute of Management Calcutta
Kolkata, India
Email: priyas@iimcal.ac.in

## Abstract

Achievement and sustenance of growth are essential themes in organizational literature. In our paper, we develop models using systems thinking approach to understand how firms achieve and sustain growth in a technology-intensive product domain. We augment these to explain the possible impact of disruptive technological innovation. We use enterprise software industry as the context where SAP has been acknowledged as the market leader. We find that product differentiation and learning effects helped SAP establish itself, and this growth was further sustained through networks and complementors. Introducing cloud computing as the disruptive innovation, we explain its impact on a firm. Analysis reveals that for the next wave of growth to occur, and to tap into newer markets, it would be imperative for SAP to create attractive cloud based offerings. We also discuss how the model can be enhanced by considering competitor dynamics.

## 1 Introduction

Firm growth in high technology industries has been an intense area of research especially in the context of increasing servitization of technology product firms (Baines et al. 2009) (Han et al. 2013). In the Enterprise Resource Planning (ERP) software industry, for instance, firm growth has been traditionally fuelled by product differentiation quite akin to physical product firms (Shang and Seddon 2000). However, such growth has been dominantly sustained through learning effects more common in innovative firms. The sustenance of such growth and the response of firms to disruptive innovations especially in high-technology industries is an area of on-going discussion in strategy, organization behaviour as well as information systems literature.

In our paper, we attempt to synthesize an understanding of drivers of firm growth in technology intensive industries, specifically in the context of the Enterprise Resource Planning (ERP) software industry. From the ERP vendor perspective, the market is dominated by SAP which possesses the greatest share of the enterprise software market worldwide. We therefore use the case of SAP to understand the notion of growth and the impact of disruptive innovation on firm dominance in an industry. In this backdrop, we treat cloud computing as the disruptive innovation encountered by existing players in this industry.

Cloud computing is a new paradigm in computing (Pallis 2010) (Zhang et al. 2010) although the concept of time sharing, virtualization and service oriented architecture have their roots in earlier technological environments (Hayes 2008). It enables the provision of on demand computing resources like infrastructure, platforms, and software as pay-per-use services, thereby reducing user firms' need to own and manage them individually. In doing so, it destabilizes the prevailing physical product delivery based market and creates a new one (Sultan 2014).

The effect of this disruption on user firms in several industries like education (Sultan 2010) and manufacturing (Zhang et al. 2014) has received much attention in literature. However, the implications of cloud computing for software vendor firms are not clearly understood. Using a systems thinking approach, we identify the main mechanisms that result in SAP's continued position as a market leader, and the possible impact of cloud computing on this position, and the industry as a whole.

This paper is organized as follows. We first briefly discuss the modelling approach used in the paper. Subsequently we develop a model to explain the growth and dominance of SAP in the enterprise software industry. We further develop this model to demonstrate the possible impact of technological innovation in the form of cloud computing. Finally we discuss how the model can be further enriched to include competitor dynamics, so that the impact of cloud computing on industry structure can also be inferred.

## 2 The Modelling Approach

The growth of a firm involves a complex interplay between various factors along with interesting feedback effects and significant delays. A systems thinking approach (Senge 1995) (Sterman 2000) allows us to delve deep into these underlying mechanisms and understand the dynamics of the system, based on the premise that behaviour of a system is a result of the behaviour of the constituent parts and their interactions. These dynamics arise through two types of loops, namely, reinforcing and balancing. A reinforcing loop is a positive loop where a given change amplifies the original change, whereas a balancing loop is a negative loop where dampening occurs. A system may consist of any number of these loops, and the final state of the system can be determined by analysing the various loop dominances (Sterman 2000).

In this paper, we use the systems thinking approach to develop causal loop structures that explain the various mechanisms by which the growth of a firm like SAP occurred, and the impact of technological innovation, in the form of cloud computing. To build our model of firm growth, we draw on published academic documents on SAP and use them as the source of secondary information on SAP. For the purpose of model building for this paper, we relied heavily on the literature summarized in table 1 below. We have also used information about the ERP industry and SAP in particular, published in popular business magazines to help us build the models in the paper. These are not mentioned in the table below.

| Author(s) | Contribution to our Model |
|---|---|
| Rashid et al. (2002) | Evolution of SAP in the context of the industry |
| Benders et al. (2006) | Role of network effects in SAP's growth |
| Lehrer (2006) | Description of SAP's initial growth, specifically product differentiation and learning effects |
| Jacobs and Weston Jr. (2007) | Evolution of SAP in the context of the industry |
| Leimbach (2008) | Description of SAP's initial growth via learning and complementor effects |
| Huang et al. (2012) | Role of network effects in SAP's growth, specifically the significance of knowledge spillovers |

*Table 1 List of Papers Treated as Main Secondary Sources of Information on SAP*

We combine this with relevant literature on disruptive innovation, cloud computing, servitization and enterprise software to develop our model. In the following section, we provide a brief background about the enterprise software industry, SAP, and cloud computing.

## 3 The ERP Industry, SAP and Cloud Computing

ERPs are highly complex information systems which offer the dual benefits of providing a unified enterprise view of the business processes in the organization and a repository of all transactions of the organization (Umble et al. 2003). They have their roots in Materials Requirements Planning (MRP) software which was developed in 1960s to cater to the increasing need to manage inventory in manufacturing companies. Progressively, financial management components were also added and so on, eventually taking the shape of the present day ERP systems. SAP has been instrumental in the evolution of ERP systems.

As organizations moved from functional to process based information systems, the adoption of ERP systems increased, making them one of the most widespread information systems (Al-Mashari 2002). External events, like Y2K compliance concerns at the turn of the millennium, also served to hasten the growth of the enterprise software industry (Jacobs and Weston Jr. 2007). ERP systems became must-have software for large enterprises. At the same time, several smaller enterprise software vendors that catered to niche markets also emerged, for example SalesForce.com which offered only CRM software. Yet, the larger firms continued to hold fort in terms of market share of ERP systems as integrated platforms. Few of the large firms that played a central role in the evolution of the enterprise industry were established in the 1970s and continue to exist today although as consolidated players. Oracle Corporation was set up in 1977, so was J.D. Edwards. J.D. Edwards was acquired by PeopleSoft which in turn was acquired by Oracle. The Baan Corporation was established in 1978 and was subsequently acquired by Infor Global Solutions in 2003. These firms continue to operate today, some continuing to provide products and services quite similar to their initial offerings, quite akin to SAP.

SAP AG was founded in Germany in 1972 by five former IBM employees with the vision of building standard application software for real time data processing. SAP stood for Systems Analysis and Program Development. By 1975, their clients could perform purchasing, inventory management and invoice verification, and link this data directly to financial accounting. In 1979, SAP R/2, an enterprise planning tool for mainframe computers, was released. SAP AG went public in 1988, by which time SAP R/2 sales had stabilized. It relied on customers for joint development of products. In 1992, after running successful pilot implementations for a few customers, SAP R/3 was released as a client-server application. In 2000, SAP became the world's third-largest independent software vendor. During the course of this journey, SAP also acquired several companies that had offerings complimentary to SAP's products (SAP 2015). The next technological wave in computing is cloud computing. As our model deals with the impact of cloud computing, we discuss this technological wave in more detail in the following section.

### 3.1 Cloud as a New Computing Paradigm

Cloud computing, as defined by the National Institute of Standards and Technology, is a model for "enabling ubiquitous, convenient, on-demand network access to a shared pool of configurable computing resources (e.g., networks, servers, storage, applications, and services) that can be rapidly provisioned and released with minimal management effort or service provider interaction" (Mell and Grance 2011). Five essential characteristics that define cloud computing are on-demand self-service, broad network access, resource pooling, rapid elasticity and measured service.

Nicholas Carr, in his much debated article, announced the end of corporate computing (Carr 2005). His observation was that information technology was becoming a utility, to be purchased from service providers and not owned and controlled like an asset. Buyya et al. (2009) state that computing is the fifth utility, after water, electricity, gas, and telephony, and that cloud computing is one of the technologies enabling this transformation. Cloud computing is compared to condominiums, where sharing of services allows users to reap economic benefits in terms of convenience and cost savings (Helland 2013). Due to these advantages, as Carr predicted, enterprises are moving towards utility computing. There is widespread belief, in both academia and practice that the geography of computation is shifting from the local computers to the cloud (Hayes 2008). According to Gartner's report on emerging technologies, cloud computing will reach the plateau of productivity, that is, it will become adopted in the mainstream, in the next two to five years (Gartner Inc. 2014).

Based on the service model, it could be called Software as a Service (SaaS), Platform as a Service (PaaS) or Infrastructure as a Service (IaaS). SaaS is the capability provided to the consumer to use the provider's applications running on a cloud infrastructure (Mell and Grance 2011). The underlying cloud infrastructure, except maybe some limited user-specific application configuration settings, is completely managed and controlled by the provider. In PaaS, the consumer can deploy onto the cloud infrastructure consumer-created or acquired applications created using programming languages, libraries, services, and tools supported by the provider. The provider manages and controls the underlying cloud infrastructure, but the control over the deployed applications remains with the consumer. In IaaS, the consumer is provided processing, storage, network, and other fundamental computing resources where it is able to deploy and run its own platforms and applications. The provider manages and controls the underlying cloud infrastructure but not operating systems, storage, and deployed applications. All these are usually in exchange for a monthly or annual subscription fee, and consumers have the option of scaling up or down, depending on their requirements, hence they are also known as pay per use, or pay as you go models.

In this paper, we focus our analysis on the SaaS service model, where the cloud provider controls and manages all the elements of cloud computing. SaaS and its enabler, cloud computing, are becoming new platforms for enterprise and personal computing (Cusumano 2010). One key difference between the traditional software model, and the SaaS model is the service property (Stuckenberg et al. 2011).

Services have four unique characteristics, namely, intangibility, heterogeneity, inseparability of production and consumption, and perishability (Regan 1963) (Zeithaml et al. 1985). Together, these are known in services literature as IHIP characteristics. Servitization denotes the shift of offerings from products to product service systems. It entails the provision of services by conventional product firms, and the development of related capabilities. Since the term was first coined in 1988 by Vandermerwe and Rada, servitization has been a widely researched area (Baines et al. 2009), especially in the fields of manufacturing and marketing. However, this phenomenon is also observed in other industries and fields, for example, in the software industry, enabled by cloud computing. In its strictest sense, in a cloud based SaaS application, only a single version of the software is made available to all the users. This requires task disaggregation facilitated by standardization and modularity of the software and the service delivery process (Susarla et al. 2010), making the development and distribution of SaaS applications easier than in the traditional on premise scenario. Hence, the emergence of the cloud based SaaS model impacts the functioning of established ERP vendors, and also reduces the barriers to new vendor entry. In the next section, we discuss our systems thinking model that explains SAP's growth, and subsequently the impact of cloud computing on this model.

# 4 The Growth Story

A dominant firm is one which accounts for a significant share of a given market and has a significantly larger market share than its next largest rival (Khemani and Shapiro 1993). In the enterprise software industry, in keeping with previous trends, SAP remained the market leader in 2012 capturing 24% of the worldwide market share, followed by Oracle (12%) (Gartner Inc. 2013). In this section, we use a systems thinking model to present a plausible explanation for how SAP was able to attain and maintain this leading position in the enterprise software industry. We do this in a sequential manner, first we develop a model to explain the initial growth of SAP based on product differentiation and learning effects, and then identify additional reinforcing loops, namely network and complementor effects that strengthened this growth.

## 4.1 Product Differentiation and Learning Effects

SAP's R/3, its most successful ERP product, was the first to cater to the need for a general-purpose technology spanning and integrating all the functions of medium-to-large companies (Lehrer 2006). Due to this, its *Software Product Attractiveness* was high, initiating the reinforcing loop of product differentiation, via increasing *Industry Demand* for SAP's product offerings, making it further successful. Greater *Sales* and thereby *Revenue* allowed SAP to invest more in *Software Product Features* to make its offering more attractive to potential clients.

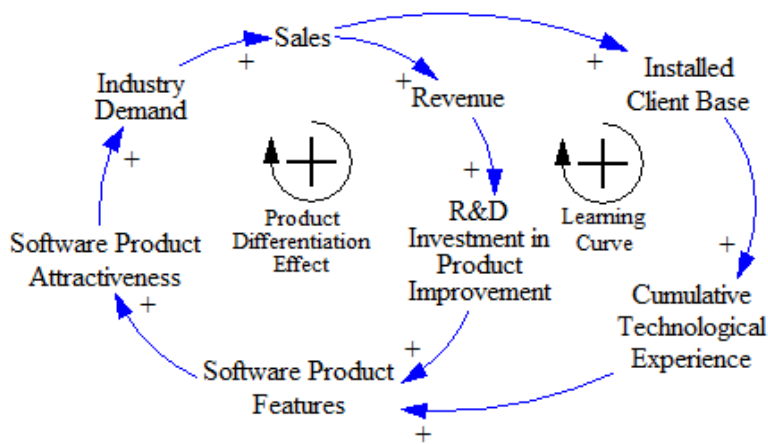

*Figure 1 Product Differentiation and Learning Effects*

As its *Installed Client Base* increased, SAP was able to utilise its *Cumulative Technological Experience* to provide more *Software Product Features*. For instance, SAP's product development was enriched by co-development with customers, especially with select pilot customers. These pilot customers include ICI Germany, SAP's first client, and John Deere, a client which prompted SAP to translate its product to English and French so that it could go for a pan-European implementation (Leimbach 2008). By continuously incorporating the needs of new clients and the inputs of the *Installed Client Base*, thereby expanding the options in its offering, and allowing clients to choose among these options, it was able to make its product more generic and hence increase *Software Product Attractiveness* (Lehrer 2006). These options eventually came to represent the best practices in the respective industries, causing firms to adopt SAP to conform to these standards, and SAP became a way of doing business, rather than just a software package (Benders et al. 2006).

Hence, the initial growth of SAP came from its addressing a market need, and sustaining that by continually strengthening its product by incremental innovations leading to a better bouquet of *Software Product Features*. However, this only partially explains the SAP growth story. The next wave of growth for SAP was based on the network and complementor effects, as explained in the next section.

## 4.2 Network and Complementor Effects

Network effect refers to the phenomenon where a good or service becomes more valuable as more people use it. According to a survey, "following the trend" was one of the top reasons influencing the choice of an ERP offering (Caldas and Wood 1999). In the Dutch energy sector, for example, the fact that several competitors had implemented SAP's ERP prompted the rest to follow suit, making it the local standard in the industry (Benders et al. 2006). The SAP Community Network was set up by SAP as a platform for interaction between SAP's users, developers, consultants, and students. This enabled them to discuss, solve each other's issues and think of new ideas (Huang et al. 2012). Increasing *Sales* thus meant that as the *Installed Client Base* increases, network effect set in, causing more firms to implement ERP, increasing *Industry Demand* for the product.

Traditionally, ERP software was implemented on premise, and is known to be a complex, expensive and time consuming endeavour, as it affects the processes and functioning of the entire enterprise (Rashid et al. 2002). The complexity of the implementation depends on the extent of customization desired by the client, among other factors. License, implementation and maintenance fees comprise the bulk of on premise ERP software costs. Even after the initial implementation is over, the buyers need to actively deal with regular improvements provided in the form of newer versions (Kremers and van Dissel 2000). Due to these complexities, this process requires prior expertise, which is often provided by trained consultants and IT service providers who help in the consultation, initial implementation and subsequent maintenance.

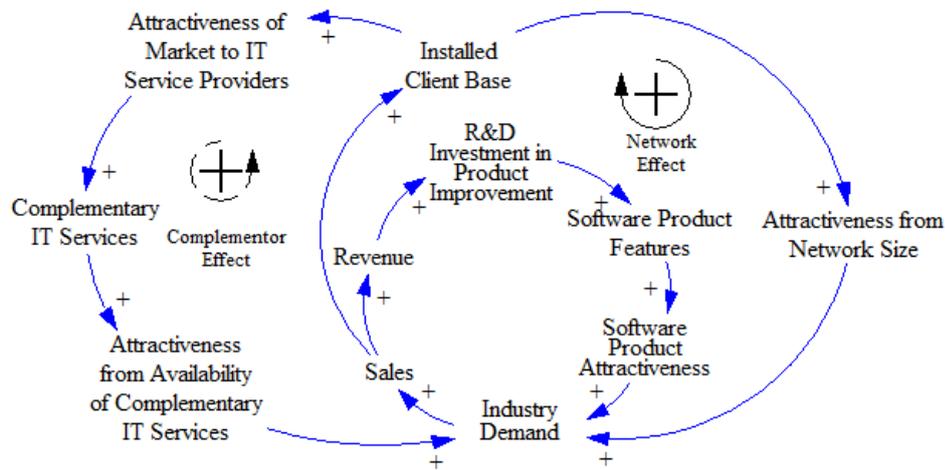

*Figure 2 Network and Complementor Effects*

Once its international expansion began, SAP had two choices, either to grow internally to support global implementations and dilute its focus on product development, or allow partners to take up reselling, implementation, maintenance, etc. SAP chose the latter route, and its partner strategy remains a focus area for the company till date (Woods 2011). It not only actively reached out to partners, who themselves were interested due to the immense growth potential shown by SAP, but also opened training schools to groom SAP consultants (Leimbach 2008). This facilitated the creation of a rich and diverse ecosystem of complementors. Complementors are organizations that increase the value of each other's offerings and the size of the total market (Yoffie and Kwak 2006).

The growth in the *Installed Client Base* thereby also had the effect of making the market attractive to complementors, namely consultants and IT service providers, triggering the proliferation of *Complementary IT Services*. For instance, in the 1990s, SAP consultancy was a highly sought after profession, so much so that R/3 consultants were generally the highest paid among consultants (Lehrer 2006). The easy availability of *Complementary IT Services* in turn enhanced *Industry Demand*, setting the reinforcing complementor effect in motion.

Thus, the existence of several positive feedback loops allowed SAP's growth to be continuously strengthened. Once SAP achieved dominance in the enterprise software industry through this growth, it was difficult for other firms to break in. However, the technological change that set in through the advent of cloud computing mandates a shift in the industry focus. In the next section, we use the causal loop structures developed so far to understand how this may play out.

## 5 Newness and Disruptive Innovation Effects

Cloud computing, as a disruptive technology, brings in a mix of prospects and concerns for both providers and consumers (Hayes 2008). Opportunities for firms, especially Small and Medium Enterprises (SMEs) to invest in enterprise systems are likely to grow in the cloud environment given the low cost, simplicity and ease of access. On the other hand, consumer concerns such as stability, privacy and dependency may reduce the potential move by user firms to the cloud.

On the provider front, cloud offers enterprise software vendors the opportunity to innovate on features drawing from both an existing set of on premise versions of their software and creating features suited to the cloud platform. Although cloud computing enhances the potential customer base for vendors by making it an attractive alternative to small enterprises, they also face risks due to demand uncertainty, inter-country data jurisdiction restrictions and a significant increase in demand for real-time and high quality customer service. The pervasiveness of service oriented architectures, especially cloud computing, is more likely to result in blurred boundaries between corporate IT and consumer IT (Fielt et al. 2013). Such overlaps are likely to impact both providers and consumers alike. Some of these are summarized in table 2 below.

|  | **Opportunities** | **Concerns** |
|---|---|---|
| **User firms** | Lower total cost of ownership (upfront and operating costs) | Security |
|  | Flexibility | Vendor lock-in |
|  | Scalability | Integration with other systems and services |
|  | Faster time to value | Business continuity |
|  | Always on the latest software release | Limited customization |
|  | Accessibility – ease of use | Jurisdiction limitations |
| **Established vendor firms** | New markets | New technology |
|  | Changed revenue streams | Uncertainty in demand |
|  | No need to maintain multiple versions | Real time customer interface |

*Table 2 Opportunities and Concerns of Cloud Computing for User and Established Vendor Firms*

Using the model developed in section 4 (figures 1 and 2) as the base, we introduce the advent of cloud computing as a game changer into the causal loop structure. Although we use the model to understand the implications for SAP, given the futuristic tone, we first present a generic explanation for firms in the enterprise software market and their response to the possible shift to cloud.

In their study of eight system development organizations, Lyytinen and Rose (2003) found that the adoption of internet computing radically impacted their IT innovation. Internet enablement of organizations, that is, the use digital networks to create value, has become imperative for all organizations (Wheeler 2002). Wheeler develops a theory of how this can be achieved through the sequential development of four capabilities, namely, choosing emerging/enabling information technologies, matching with economic opportunities, executing business innovation for growth and finally assessing customer value. However, due to path dependence, the firm's locus of search will be limited to areas where it has past experience (Chesbrough and Rosenbloom 2002) (Zahra and George 2002).

Moreover, in a relatively stable environment, as a firm's *Cumulative Technological Experience* increases, the *Pressure to Innovate* reduces. The innovation encompassed in this variable is only radical, disruptive innovation. Cloud computing, being a new technology, increases the *Pressure to Innovate* (Magnusson et al. 2012), compelling the firm to conduct *Software Product Innovation* to create cloud based product offerings, which increase *Software Product Attractiveness (Cloud Based)*, but also result in *Adoption Concerns* due to the newness of the technology and product.

In this model, the *Software Product Attractiveness* variable has been split into two, namely *Software Product Attractiveness (On Premise)* and *Software Product Attractiveness (Cloud Based)*. Since the cloud based ERP also draws from the existing on premise product, *Software Product Attractiveness (Cloud Based)* incorporates the attractiveness due to the existing *Software Product Features*. However, certain other factors also contribute to *Software Product Attractiveness (Cloud Based)*.

Cloud computing reduces the client firms need for *In-house Infrastructure Investment*, leading to lower *Total Cost of Ownership,* which increases *Software Product Attractiveness (Cloud Based)* (Utzig et al. 2013). Several researchers and industry experts have drawn comparison between ERP offered on premise, and cloud based ERP (Johansson and Ruivo 2013; Mattison and Raj 2012). These include both advantages of cloud based ERP, like increased scalability and mobility and reduced implementation times, and challenges like data security issues and internet dependency.

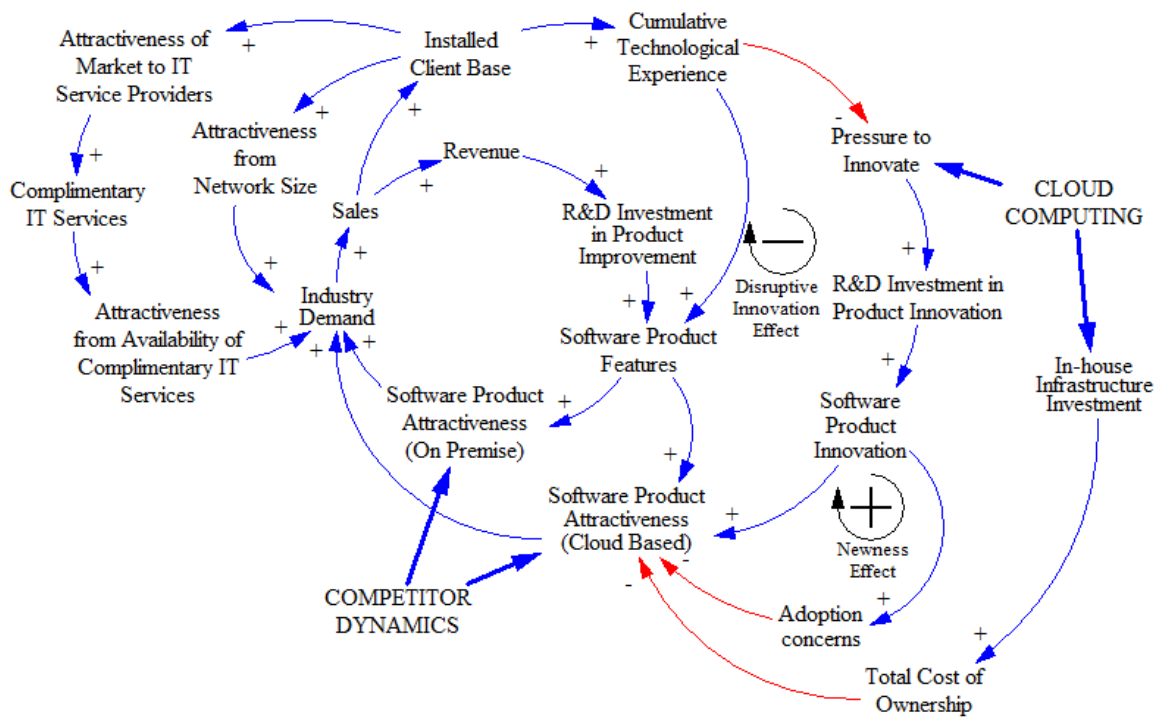

*Figure 3 Impact of Cloud Computing on Firm Growth*

It is important to note here that firm size, particularly the distinction between SMEs and large companies, has been an important aspect of enterprise software research. Due to the cost and complexity associated with ERP implementation, it is typically large firms that purchased and implemented ERP. However, as ERP adoptions in large firms began to saturate, ERP vendors began targeting SMEs (Haddara and Zach 2011). Cloud based ERPs, due to the possible reduction of both cost and complexity; make ERPs more attractive to SMEs. The scale of the SME opportunity can be gauged from the fact that, according to a World Bank report, there are there are 125 million formal MSMEs in the 132 economies covered (Kushnir et al. 2010).

Hence, by responding to the *Pressure to Innovate* by *Software Product Innovation* to develop an attractive cloud based ERP, SAP can usher in a new wave of growth, by targeting the previously under-attended category of firms, namely SMEs and therefore, increase the *Industry Demand*.

Towards this, SAP launched Business ByDesign, its foray into cloud based ERP, in 2007. Realizing that cloud computing was becoming increasingly popular, it also acquired several companies that had offerings complimentary to its products, including Ariba and SuccessFactors for cloud computing, and Sybase, for its capabilities in the mobile devices domain. However, as discussed in section 3.1, cloud based SaaS ERP lowers the barriers to entry for new vendors, by enabling task disaggregation. ERP vendors that cater to niche requirements, like Salesforce.com and cloud first ERP vendors like NetSuite have made a mark in the enterprise software market. Cloud computing, therefore, necessitates a look at how competitor dynamics impacts our model.

Competitor dynamics has been considered as an external variable in our model. The attractiveness of the competitor firms' software products influences the relative attractiveness of a firm's own product, both on premise and cloud based. Several researchers assert that the firms that use disruptive technology successfully are not firms that occupy the dominant position in an industry, but those that are struggling for survival (Christensen 1997) (Henderson and Clark 1990) (Tushman and O'Reilly 1996). Looking for avenues to improve their performance, these firms generally innovate with the latest available technology. Dominant, incumbent firms, on the other hand, are either blind-sided by their continued success, or unable to act fast enough when disruptive technology emerges (Walsh et al. 2002). Hence, organizational inertia often prevents established firms from capitalising on disruptive technological innovation (Tripsas and Gavetti 2000). By sensing and monitoring its environment the incumbent, here SAP, can perceive changes faster, and hence react before the disruptive innovation effect enables competitors firms to emerge. The incumbent's response would be increasing its *R&D Investment* in *Product Innovation*, thereby giving it ample resources to create products capitalizing on the new technology, and hence cater to the *Industry Demand* and sustain its growth.

This can be explained as follows by our model. *Software Product Innovation* in response to cloud computing also comes with concerns, which can be broadly categorised into two types, namely those of the cloud computing technology itself, and those that are specific to a vendor's cloud offering. Although the former category affects all enterprise software vendors, the latter is under the control of the vendor firm. Thus, when we incorporate competitor dynamics into the model, we can see how this feeds into *Software Product Attractiveness* of various ERP products available. If the response of the vendor firm, here SAP, to cloud computing as a technological change is not rapid or appropriate enough, there is a possibility of change in industry structure and dynamics.

A caveat for the future is the possible cannibalization between SAP's on premise and cloud based ERPs. This can also be analysed by enhancing our model by incorporating the interplay between *Software Product Attractiveness (Cloud Based)* and *Software Product Attractiveness (On Premise)*, instead of both contributing to a combined *Industry Demand* for SAP's ERP offerings. Hence, it would be important for SAP to not only define its cloud strategy so as to be able to capitalize on it, and transform it into its next growth opportunity, but also differentiate it from its existing on premise offerings.

# 6 Conclusion

The growth story of the enterprise software industry has also been SAP's growth story. Having been founded in the 1970s, when the industry was just taking shape, SAP achieved and maintained the position of a market leader. Our paper has used the secondary information available on SAP to model this growth, using the systems thinking approach. This allows us to look at firm growth from the perspective of how continued advantage can be built through the use of reinforcing loops, and how responses to technological changes can disturb the system.

SAP began its growth trajectory by creating a product that catered to a market need. It was able to establish itself by continually differentiating its product by investing in R&D, and learning from its increasing customer base, capturing its experience and knowledge from its installations, and building on it. As its pace of growth increased, and to ensure it is not limited by capacity and geography, it actively engaged partners that provided support to customers for ERP selection, implementation and maintenance. SAP thereby gained access to newer markets at a faster rate, and at the same time was able to focus on product development.

However, new technologies typically cause a disruption in the industry. Cloud computing is one such technology that is leading to structural changes in the software industry via newness and disruptive innovation effects. By enabling the servitization of products, cloud computing allows smaller vendor firms a fair chance at succeeding in the enterprise software market in several ways. Examples of such firms include NetSuite and Epicor. The subscription fees model provides a continuous revenue stream for these firms, and is a marked change from the dependence on infrequent but large licence and maintenance fees. At the same time, cloud computing expands the enterprise software market by lowering initial fixed costs, and providing advantages like scalability, thereby increasing accessibility to small and medium enterprises, a fast growing and previously under-addressed segment. Hence, cloud as a service computing platform demands further research in the IS discipline in order to understand its impact and implications while evolving conceptual frameworks, models and theories (Fielt et al. 2013) that capture the nuances of organizational technology choices from the consumer perspective along with strategic stance and product choices from the vendor perspective.

# 7 Limitations and Future Research

In this paper, we have sequentially developed a model to explain firm growth and sustenance in the enterprise software industry based on secondary information of SAP. Although there exists sufficient literature that documents SAP's growth story against the backdrop of the global software industry, our model could have been further strengthened through primary data gathered in the form of interviews with experts. Secondly, although our model is based chiefly on the evolution and growth of a single firm, namely SAP, it does have the potential to be used as a generic model of firm growth and sustenance, especially in industries with similar structures. This can be done by validating this model with the growth stories of other firms, and we hope to do so in subsequent extensions of this research. Thirdly, while we do consider the impact of competitor dynamics on the attractiveness of a firm's product, and how cloud computing is enabling new entrants, we do not delve into how these dynamics emerge. A deeper understanding of competitor dynamics is hence required, and would help enrich the model further.

## 8 References

Al-Mashari, M. 2002. "Enterprise resource planning (ERP) systems: a research agenda," *Industrial Management & Data Systems*, pp. 165–170 (doi: 10.1108/02635570210421354).

Baines, T. S., Lightfoot, H. W., Benedettini, O., and Kay, J. M. 2009. "The servitization of manufacturing: A review of literature and reflection on future challenges," *Journal of Manufacturing Technology Management* (20:5), pp. 547–567 (doi: 10.1108/17410380910960984).

Benders, J., Batenburg, R., and Van Der Blonk, H. 2006. "Sticking to standards; Technical and other isomorphic pressures in deploying ERP-systems," *Information and Management* (43:2), pp. 194–203 (doi: 10.1016/j.im.2005.06.002).

Buyya, R., Yeo, C. S., Srikumar, V., Broberg, J., and Brandic, I. 2009. "Cloud computing and emerging IT platforms: Vision, hype, and reality for delivering computing as the 5th utility," *Future Generation Computer Systems* (25:6), pp. 599–616 (available at http://www.sciencedirect.com/science/article/pii/S0167739X08001957).

Caldas, M. P., and Wood, T. J. 1999. "How consultants can help organizations survive the ERP frenzy," in *Annual meeting of the Academy of Management*, Chicago (available at http://citeseerx.ist.psu.edu/viewdoc/download?doi=10.1.1.197.1330&rep=rep1&type=pdf).

Carr, N. G. 2005. "The End of Corporate Computing," *MIT Sloan Management Review* (46:3), pp. 62–73 (doi: 821058931).

Chesbrough, H., and Rosenbloom, R. S. 2002. "The role of the business model in capturing value from innovation: evidence from Xerox Corporation's technology spin-off companies," *Industrial and Corporate Change* (11:3), pp. 529–555 (doi: 10.1093/icc/11.3.529).

Cusumano, M. 2010. "Cloud Computing and SaaS as New Computing Platforms," *Communications of the ACM* (53:4), pp. 27–29.

Fielt, E., Böhmann, T., Korthaus, A., Conger, S., and Gable, G. 2013. "Service Management and Engineering in Information Systems Research," *The Journal of Strategic Information Systems* (22:1), pp. 46–50 (doi: 10.1016/j.jsis.2013.01.001).

Gartner Inc. 2013. "Market Share Analysis: ERP Software, Worldwide," (available at https://www.gartner.com/doc/2477517/market-share-analysis-erp-software).

Gartner Inc. 2014. "Gartner's Hype Cycle Special Report for 2014," (available at http://www.gartner.com/technology/research/hype-cycles/).

Haddara, M., and Zach, O. 2011. "ERP Systems in SMEs: A Literature Review," in *44th Hawaii International Conference on System Sciences*, Hawaii: IEEE, January, pp. 1–10 (doi: 10.1109/HICSS.2011.191).

Han, S., Kuruzovich, J., and Ravichandran, T. 2013. "Service Expansion of Product Firms in the Information Technology Industry: An Empirical Study," *Journal of Management Information Systems* (29:4), pp. 127–158 (doi: 10.2753/MIS0742-1222290405).

Hayes, B. 2008. "Cloud computing," *Communications of the ACM* (51:7), p. 9 (doi: 10.1145/1364782.1364786).

Helland, P. 2013. "Condos and clouds," *Communications of the ACM* (56:1), p. 50 (doi: 10.1145/2398356.2398374).

Huang, P., Ceccagnoli, M., Forman, C., and Wu, D. 2012. "IT Knowledge Spillovers and Productivity: Evidence from Enterprise Software," in *Workshop on Information Systems and Economics (WISE)*, Orlando, pp. 0–41.

Jacobs, F. R., and Weston Jr., F. C. "Ted." 2007. "Enterprise Resource Planning (ERP)-A Brief History," *Journal of Operations Management* (25:2), pp. 357–363 (doi: 10.1016/j.jom.2006.11.005).

Johansson, B., and Ruivo, P. 2013. "Exploring Factors for Adopting ERP as SaaS," *Procedia Technology* (9), Elsevier B.V., pp. 94–99 (doi: 10.1016/j.protcy.2013.12.010).

Khemani, R. S., and Shapiro, D. . 1993. *Glossary of industrial organisation economics and competition law*, Paris: Organisation for Economic Co-operation and Development (OECD).

Kremers, M., and van Dissel, H. 2000. "Enterprise resource planning: ERP system migrations," *Communications of the ACM* (43:4), pp. 53–56 (doi: 10.1145/332051.332072).

Kushnir, K., Mirmulstein, M. L., and Ramalho, R. 2010. "Micro, Small, and Medium Enterprises Around the World: How Many Are There, and What Affects the Count?," *World Bank, IFC* (available at http://www.ifc.org/wps/wcm/connect/9ae1dd80495860d6a482b519583b6d16/MSME-CI-AnalysisNote.pdf?MOD=AJPERES).

Lehrer, M. 2006. "Two types of organisational modularity: SAP, ERP product architecture and the German tipping point in the make/buy decision for IT services," in *Knowledge Intensive Business Services: Organizational Forms and National Institutions* M. Miozzo and D. Grimshaw (eds.), Cheltenham: Edward Elgar Publishing Limited, pp. 187–204.

Leimbach, T. 2008. "The SAP Story: Evolution of SAP within the German Software Industry," *IEEE Annals of the History of Computing* (30:4), pp. 60–76 (doi: 10.1109/MAHC.2008.75).

Lyytinen, K., and Rose, G. M. 2003. "The disruptive nature of information technology innovations: the case of internet computing in systems development organizations," *MIS Quarterly* (27:4), pp. 557–596.

Magnusson, J., Enquist, H., Juell-Skielse, G., and Uppström, E. 2012. "Incumbents and Challengers: Conflicting Institutional Logics in SaaS ERP Business Models," *Journal of Service Science and Management* (05:01), pp. 69–76 (doi: 10.4236/jssm.2012.51009).

Mattison, B. J. B., and Raj, S. 2012. "Key questions every IT and business executive should ask about cloud computing and ERP.,"

Mell, P., and Grance, T. 2011. "The NIST Definition of Cloud Computing Recommendations of the National Institute of Standards and Technology," *National Institute of Standards and Technology, Information Technology Laboratory* (145), p. 7 (doi: 10.1136/emj.2010.096966).

Pallis, G. 2010. "Cloud Computing: The New Frontier of Internet Computing," *IEEE Internet Computing* (14:5), pp. 70–73 (doi: 10.1109/MIC.2010.113).

Rashid, M. A., Hossain, L., and Patrick, J. D. 2002. "The Evolution of ERP Systems: A Historical Perspective," in *Enterprise Resource Planning: Global Opportunities and Challenges*M. Khosrowpour, J. Travers, M. Rossi, and B. Arneson (eds.), Hershey: Idea Group Publishing, pp. 1–16 (doi: 10.4018/978-1-931777-06-3.ch001).

Regan, W. J. 1963. "The Service Revolution," *Journal of Marketing* (27:3), p. 57 (doi: 10.2307/1249437).

SAP. 2015. "A 43-year history of innovation," (available at http://www.sap.com/corporate-en/about/our-company/history/index.html).

Senge, P. 1995. *The Fifth Discipline : The Art and Practice of the Learning Organization* (1st editio.), New York, NY: Doubleday/Currency.

Shang, S., and Seddon, P. B. 2000. "A comprehensive framework for classifying the benefits of ERP systems". *AMCIS 2000 proceedings*, p. 39.

Sterman, J. 2000. *Business Dynamics: Systems Thinking and Modeling for a Complex World* (First.), New York, NY: McGraw-Hill New York.

Stuckenberg, S., Fielt, E., and Loser, T. 2011. "The impact of software-as-a-service on business models of leading software vendors: experiences from three exploratory case studies," *PACIS 2011 Proceedings*, p. 17 (available at http://eprints.qut.edu.au/43815/).

Sultan, N. 2010. "Cloud computing for education: A new dawn?," *International Journal of Information Management* (30:2), pp. 109–116 (doi: 10.1016/j.ijinfomgt.2009.09.004).

Sultan, N. 2014. "Servitization of the IT Industry: The Cloud Phenomenon," *Strategic Change* (doi: 10.1002/jsc.1983).

Susarla, A., Barua, A., and Whinston, A. B. 2010. "Multitask Agency, Modular Architecture, and Task Disaggregation in SaaS," *Journal of Management Information Systems* (26:4), pp. 87–118 (doi: 10.2753/MIS0742-1222260404).

Umble, E. J., Haft, R. R., and Umble, M. M. 2003. "Enterprise resource planning: Implementation procedures and critical success factors," *European Journal of Operational Research* (146:2), pp. 241–257 (doi: 10.1016/S0377-2217(02)00547-7).

Utzig, C., Holland, D., Horvath, M., and Manohar, M. 2013. "ERP in the cloud: Is it ready ? Are you ?," (available at http://www.strategyand.pwc.com/media/file/Strategyand_ERP-in-the-Cloud.pdf).

Wheeler, B. C. 2002. "NEBIC: A Dynamic Capabilities Theory for Assessing Net-Enablement," *Information Systems Research* (13:2), pp. 125–146 (doi: 10.1.1.199.3070).

Woods, D. 2011. "How SAP is Betting Its Growth on Partnerships," *Forbes* (available at http://www.forbes.com/sites/ciocentral/2011/09/12/how-sap-is-betting-its-growth-on-partnerships/; retrieved August 6, 2015).

Yoffie, D. B., and Kwak, M. 2006. "With friends like these: The art of managing complementors," *Harvard Business Review* (84:9), pp. 88–98 (doi: 10.1007/BF02208458).

Zahra, S., and George, G. 2002. "The Net-enabled business innovation cycle and a strategic entrepreneurship perspective on the evolution of dynamic capabilities," *Information Systems Research* (13:2), pp. 147–150 (doi: 10.1287/isre.13.2.147.90).

Zeithaml, V. A., Parasuraman, A., and Berry, L. L. 1985. "Problems and strategies in services marketing," *The Journal of Marketing* (49:2), pp. 33–46 (doi: 10.2307/1251563).

Zhang, L., Luo, Y., Tao, F., Li, B. H., Ren, L., Zhang, X., Guo, H., Cheng, Y., Hu, A., and Liu, Y. 2014. "Cloud manufacturing: a new manufacturing paradigm," *Enterprise Information Systems* (8:2), pp. 167–187 (doi: 10.1080/17517575.2012.683812).

Zhang, Q., Cheng, L., and Boutaba, R. 2010. "Cloud computing: state-of-the-art and research challenges," *Journal of Internet Services and Applications* (1:1), pp. 7–18 (doi: 10.1007/s13174-010-0007-6).